\documentclass{sigchi-ext}
\usepackage[T1]{fontenc}
\usepackage{textcomp}
\usepackage[scaled=.92]{helvet} 
\usepackage{graphicx} 
\usepackage{balance}  
\usepackage{booktabs} 
\usepackage{ccicons}  
\usepackage{ragged2e} 

\title{Wikidata: A New Paradigm of Human-Bot Collaboration?}

\numberofauthors{1}

\author{%
  \alignauthor{%
    \textbf{Alessandro Piscopo}\\
    \affaddr{University of Southampton} \\
    \affaddr{Southampton, United Kingdom} \\
    and \\
    \affaddr{The Open Data Institute} \\
    \affaddr{London, United Kingdom} \\
    \email{alessandro.piscopo@gmail.com} }}

\definecolor{linkColor}{RGB}{6,125,233}

\begin{document}


\maketitle

\RaggedRight{} 


\begin{abstract}
  Wikidata is a collaborative knowledge graph which has already drawn the attention of practitioners and researchers. It is the work of a community of volunteers, supported by policies, guidelines and automatic programs (bots) which perform a broad range of tasks, doing the lion's share of the work on the platform. In this paper, we highlight some of the most salient aspects of human-bot collaboration in Wikidata. We argue that the combination of automated and semi-automated work produces new challenges with respect to other online collaboration platforms. 
\end{abstract}

\keywords{Wikidata; bots; collaborative knowledge engineering}

\section{Introduction}
Wikidata is a relatively young project, yet it has already drawn great attention from practitioners and researchers. It is a collaborative knowledge graph---a knowledge base that describes real world entities and the relationships that occur among them, organised in a graph~\cite{DBLP:journals/semweb/Paulheim17}. 
Several features make Wikidata worthy of interest. Since its inception in $2012$ it has already gathered a community of $\sim18$k monthly active users, who have built a graph that covers facts about around $45$M entities. In relation to CSCW, Wikidata has been considered by some researchers as a new time of platform, at the intersection of peer-production systems and collaborative ontology development projects~\cite{DBLP:conf/wikis/Muller-BirnKLL15}.\\  
The steep growth of Wikidata can be attributed in a large part to the work of bots, pieces of software programmed to perform a range of tasks, among which importing new data from different sources. Especially in the early years of Wikidata's life, bots' contributions have boosted the growth of the graph, adding a large amount of facts. Whereas this has allowed
Wikidata to reach a size large enough for users to build upon it and produce a usable knowledge source, it has also posed some challenges, regarding the quality of automated work and its effects on the community. Although some of these challenges have been outlined by prior work---e.g. the difficulty for editors to control the quality of bot-generated data~\cite{DBLP:conf/semweb/PiscopoKPS17}---they have never been clearly identified. This is the aim of this paper. Understanding what these challenges are and addressing them is key to ensure the quality and the future sustainability of Wikidata. 

\marginpar{%
  \vspace{-210pt} \fbox{%
    \begin{minipage}{0.925\marginparwidth}
      \textbf{Data} \\
      The Wikimedia Foundation releases dumps containing the whole history of Wikidata revisions. These contain not only the data within each page (item and properties) included, but also the related metadata, e.g revision id, username of the author of the revision, and a timestamp. All figures in this paper have been extracted by dumps updated to 1st October $2017$.
\end{minipage}}\label{sec:sidebar} }

\begin{margintable}
    \centering
    \caption{Number of users and edits per type. Please note that anonymous human users refer to unique IPs, but nothing prevents users from connecting from different IPs.}
    \label{tab:user_types}
    \setlength{\tabcolsep}{2.1pt}
    \begin{tabular}{p{1.4cm}  c  c }
         &  {\small \textbf{\# users}} & {\small \textbf{\# edits}} \\
         \toprule
         {\small Bots} & {\small$407$} & {\small$384,660,528$} \\ 
        {\small Registered humans} & {\small$190,765$} & {\small$171,824,150$} \\
        {\small Anonymous humans} & {\small$548,956$} & {\small$2,329,109$} \\
        \bottomrule
    \end{tabular}
\end{margintable}

\section{The Knowledge Graph}
\textit{Items} and \textit{properties} are the building blocks of Wikidata's knowledge: items represent instances of concrete or abstract entities---e.g. Lou Reed or New York---as well as classes---e.g. the class of all musicians. Properties are used to state facts about items, such as \textit{Lou Reed}--\textit{place of birth}--\textit{New York}. 
\textit{Statements} assert facts about items and properties. Their nucleus is a \textit{claim}, a property-value pair whose value can be either an item or a literal. Claims can be enriched through \textit{qualifiers} and \textit{references}. Qualifiers add contextual information (e.g. specifying a limitation in the validity of a statement), whereas references link to a source. The knowledge graph is the set of all statements. 

\section{Editing Wikidata}
\begin{figure}
  \includegraphics[width=\columnwidth]{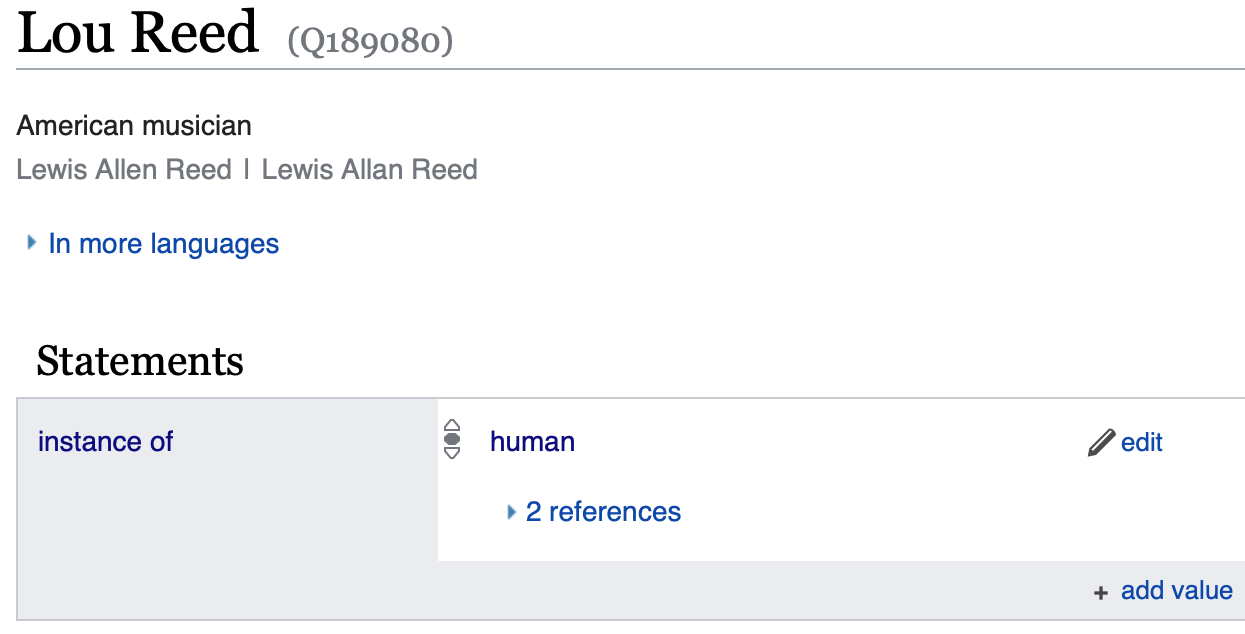}
  \caption{The web interface of a Wikidata item}~\label{fig:lou}
  \vspace{-5pt}
\end{figure}

The Wikidata community has continuously grown along the whole lifespan of the project, reaching a total of more than $190$ thousand unique registered users. Editors do not need to register to contribute and can do that also anonymously. Similarly to what observed in other online collaboration projects (e.g. Wikipedia~\cite{DBLP:conf/hicss/OrtegaGR08}), the distribution of edits is very skewed and a core of users carry out the bulk of the work~\cite{sarasua_wikidata}. This is facilitated by some tools that are peculiar to the project. Wikidata can be edited through several interfaces. The easiest one is the web interface. Every entity in Wikidata has a corresponding a web page, which can be retrieved and variously edited (Figure~\ref{fig:lou}). 
Another commonly used interface are \textbf{semi-automated editing tools}, such as \textit{QuickStatements} or \textit{The Wikidata Game}. These allow users to edit at a much higher rate than it would be possible through the web interface, e.g. QuickStatements accepts csv files with a list of statements to be added, or check the quality of suggested statements. Revisions made through these tools account for around $60\%$ of all edits made by human editors, although they are used by less than $5\%$.

\section{Bots}
\begin{figure*}[!h] 
    \centering
    \includegraphics[width=1.9\columnwidth]{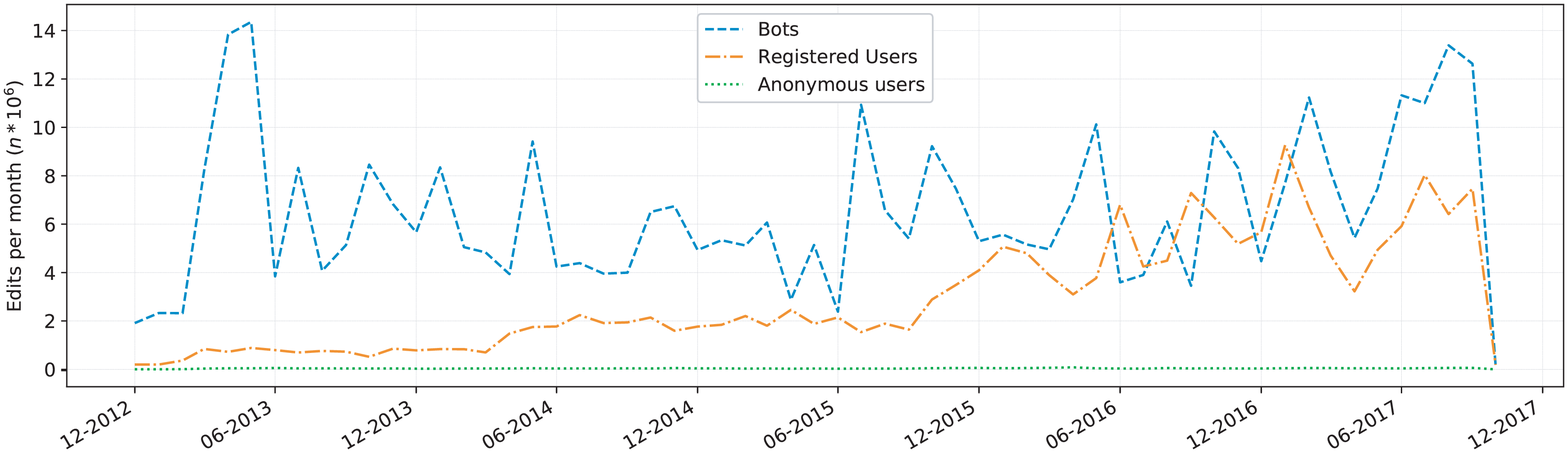}
    \caption{Number and percentage of edits per user type}
    \label{fig:user_type}
\end{figure*}

Bots carry out various types of tasks, such as editing items and properties or patrolling the graph for quality control checks. They are the authors of the majority of edits on Wikidata. Although their percentage of edits over the total has declined since the early years of the project, when they exceeded $90\%$ of all contributions~\cite{DBLP:conf/wikis/Steiner14}, they remains above $50\%$ of all revisions (Figure~\ref{fig:user_type}). 

Bots are operated by registered users, according to codified norms and policies. The first step to obtain the community approval to run a bot is to open a \textit{Request for permission}\footnote{\url{https://www.wikidata.org/wiki/Wikidata:Requests_for_permissions/Bot}, consulted on $26$ September $2018$.}, where editors must provide a detailed description of the activities that their bot is planned to do. After a test run (between $50$ and $250$ edits), other users can leave their comment and vote in favour or against the activation of the bot. Once a bot is granted the community approval, its ``owner'' must continuously check its work and, if it becomes harmful for the graph, immediately suspend it. Other users can request to withdraw the authorisation by opening a new page on a dedicated section of Wikidata. Unauthorised bots exist, but an editing cap is enforced on them. 
One of the first functions for Wikidata items was to act as inter-language hub for different language versions of Wikipedia articles~\cite{DBLP:journals/cacm/VrandecicK14}. Therefore, the first Wikidata bots harvested inter-language links from all Wikipedias, importing them over to Wikidata, where each item was connected to the corresponding articles in several language versions of the online encyclopaedia. Besides this first task, bot activities have been very focused on importing new statements and enrich the knowledge graph. Almost $90\%$ of automated editing in Wikidata concerns the addition or modification of Item statements ($58\%$) or labels/descriptions/aliases ($30\%$)~\cite{DBLP:conf/wikis/Muller-BirnKLL15}. Nonetheless, bots take roles similar to human editors in constructing Wikidata's knowledge. \cite{DBLP:conf/wikis/Muller-BirnKLL15} have shown that bots are active in five of the six user roles they identified in Wikidata on the basis of type and scope of edits, i.e.reference editor, item creator, item editor, item expert, and property editor. Other bot activities concern quality-control tasks. A bot called KrBot has regularly scanned the whole graph since April $2013$ and reported property constraints violations.

Bots play a prominent role in the production of Wikidata's knowledge, as we have seen in the previous paragraphs. This is true also with regard to other projects. For instance, bots are by all means part of the sociotechnical fabric of Wikipedia, providing an essential contribution to authoring and cleaning articles~\cite{DBLP:journals/nms/NiedererD10} and are crucial to respond timely to vandalism~\cite{DBLP:conf/wikis/GeigerH13}. Nevertheless, we argue that Wikidata constitutes a new model of human-bot collaboration, due to its combination of automated and semi-automated work, presenting new challenges. We outline these in the following:

\textbf{1.} Bots generate new content massively. However, most of this content is likely to be never seen not checked by any human user. With more than $45$M entities in the graph,large swathes of it may be never consulted by anyone. Moreover, Wikidata is often accessed through third party applications which, according to Wikidata's CC0 licence, do not need to provide attribution. The claim that ``given enough eyeballs all bugs are shallow'' may not work in this case, simply because there might not be enough eyeballs.
 
\textbf{2.} Bots import large numbers of statements from a small number of sources, leading to a lack of diversity of the knowledge they produce~\cite{DBLP:conf/semweb/PiscopoKPS17}. Combined to the fact that a restricted circle of users operate bots and that a very small core of human editors perform the majority of edits through semi-automated tool, this can be a serious threat to representing a broad range of viewpoints in Wikidata, a project that was designed having diversity in mind~\cite{DBLP:journals/cacm/VrandecicK14}.

\textbf{3.} The extensive proportion of automated and semi-automated activity, together with the fact that Wikidata is a multilingual project, may stifle user participation on the platform. Whereas an in-depth study of this aspect has not been carried out yet, it must be noted that discussion pages are seldom used (only $10,787$ item have active talk pages). Does communication between users decrease, compared to other platforms? And if so, how does that affect the point above, i.e. diversity?

\section{Conclusions}
Research has already investigated some of the effects of algorithmic work in Wikidata on data quality (e.g. \cite{DBLP:conf/semweb/PiscopoKPS17} and~\cite{DBLP:conf/socinfo/PiscopoPS17}). However, a study of the main challenges posed by bot activity in Wikidata is still missing. This paper highlights three of them, namely \textit{quality control}, \textit{lack of diversity}, and \textit{threats to user participation}. Further work should address them, in order to shed light on these aspects of algorithmic participation in online platforms and ensure the future sustainability of Wikidata.

\bibliographystyle{SIGCHI-Reference-Format}
\bibliography{workshop}


\begin{thebibliography}{00}


\ifx \showCODEN    \undefined \def \showCODEN     #1{\unskip}     \fi
\ifx \showDOI      \undefined \def \showDOI       #1{{\tt DOI:}\penalty0{#1}\ }
  \fi
\ifx \showISBNx    \undefined \def \showISBNx     #1{\unskip}     \fi
\ifx \showISBNxiii \undefined \def \showISBNxiii  #1{\unskip}     \fi
\ifx \showISSN     \undefined \def \showISSN      #1{\unskip}     \fi
\ifx \showLCCN     \undefined \def \showLCCN      #1{\unskip}     \fi
\ifx \shownote     \undefined \def \shownote      #1{#1}          \fi
\ifx \showarticletitle \undefined \def \showarticletitle #1{#1}   \fi
\ifx \showURL      \undefined \def \showURL       #1{#1}          \fi

\bibitem{DBLP:conf/wikis/GeigerH13}
{R.~Stuart Geiger} {and} {Aaron Halfaker}. 2013.
\newblock \showarticletitle{When the levee breaks: without bots, what happens
  to {W}ikipedia's quality control processes?}. In {\em OpenSym}. {ACM},
  6:1--6:6.
\newblock


\bibitem{DBLP:conf/wikis/Muller-BirnKLL15}
{Claudia M{\"{u}}ller{-}Birn}, {Benjamin Karran}, {Janette Lehmann}, {and}
  {Markus Luczak{-}R{\"{o}}sch}. 2015.
\newblock \showarticletitle{Peer-production system or collaborative ontology
  engineering effort: what is {W}ikidata?}. In {\em OpenSym}. {ACM},
  20:1--20:10.
\newblock


\bibitem{DBLP:journals/nms/NiedererD10}
{Sabine Niederer} {and} {Jos{\'{e}} van Dijck}. 2010.
\newblock \showarticletitle{Wisdom of the crowd or technicity of content?
  Wikipedia as a sociotechnical system}.
\newblock {\em New Media {\&} Society\/} {12}, 8 (2010), 1368--1387.
\newblock
\showDOI{%
\url{http://dx.doi.org/10.1177/1461444810365297}}


\bibitem{DBLP:conf/hicss/OrtegaGR08}
{Felipe Ortega}, {Jes{\'{u}}s~M. Gonz{\'{a}}lez{-}Barahona}, {and} {Gregorio
  Robles}. 2008.
\newblock \showarticletitle{On the Inequality of Contributions to Wikipedia}.
  In {\em {HICSS}}. {IEEE} Computer Society, 304.
\newblock


\bibitem{DBLP:journals/semweb/Paulheim17}
{Heiko Paulheim}. 2017.
\newblock \showarticletitle{Knowledge graph refinement: {A} survey of
  approaches and evaluation methods}.
\newblock {\em Semantic Web\/} {8}, 3 (2017), 489--508.
\newblock


\bibitem{DBLP:conf/semweb/PiscopoKPS17}
{Alessandro Piscopo}, {Lucie{-}Aim{\'{e}}e Kaffee}, {Chris Phethean}, {and}
  {Elena Simperl}. 2017a.
\newblock \showarticletitle{Provenance Information in a Collaborative Knowledge
  Graph: An Evaluation of Wikidata External References}. In {\em International
  Semantic Web Conference {(1)}} {\em (Lecture Notes in Computer Science)},
  Vol. 10587. Springer, 542--558.
\newblock


\bibitem{DBLP:conf/socinfo/PiscopoPS17}
{Alessandro Piscopo}, {Chris Phethean}, {and} {Elena Simperl}. 2017b.
\newblock \showarticletitle{What Makes a Good Collaborative Knowledge Graph:
  Group Composition and Quality in Wikidata}. In {\em SocInfo {(1)}} {\em
  (Lecture Notes in Computer Science)}, Vol. 10539. Springer, 305--322.
\newblock


\bibitem{sarasua_wikidata}
{Cristina Sarasua}, {Alessandro Checco}, {Gianluca Demartini}, {Djellel
  Difallah}, {Michael Feldman}, {and} {Lydia Pintscher}. 2018.
\newblock \showarticletitle{The Evolution of Power and Standard Wikidata
  Editors: Comparing Editing Behavior over Time to Predict Lifespan and Volume
  of Edits}.
\newblock {\em Journal of Computer Supported Cooperative Work\/} (2018),
  00--00.
\newblock


\bibitem{DBLP:conf/wikis/Steiner14}
{Thomas Steiner}. 2014.
\newblock \showarticletitle{Bots vs. {W}ikipedians, {A}nons vs. {L}ogged-Ins
  (Redux): {A} Global Study of Edit Activity on Wikipedia and Wikidata}. In
  {\em OpenSym}. {ACM}, 25:1--25:7.
\newblock


\bibitem{DBLP:journals/cacm/VrandecicK14}
{Denny Vrande\v{c}i\'{c}} {and} {Markus Kr{\"{o}}tzsch}. 2014.
\newblock \showarticletitle{Wikidata: a free collaborative knowledgebase}.
\newblock {\em Commun. {ACM}\/} {57}, 10 (2014), 78--85.
\newblock


\end{thebibliography}
\balance{} 
\end{document}